\documentclass[aps,prb,nofootinbib,longbibliography,showkeys]{revtex4-2}

\usepackage{amsfonts,amsmath,,amssymb,
		xcolor,
		dcolumn,
		framed,
		graphicx,
		lipsum,
		moreverb, multirow,
		ulem,
		verbatim
		}
\usepackage{siunitx} 
\usepackage{bm} 

\usepackage[pdftex,
            ]{hyperref}

\def\be{\begin{equation}}
\def\ee{\end{equation}}
\def\ber{\begin{eqnarray}}
\def\eer{\end{eqnarray}}

\def\bsigma{\mbox{\boldmath $\sigma$}}
\def\br{{\bf r}}

\def\bb{{\bf b}}

\def\bp{{\bf p}}

\def\bv{{\bf v}}
\def\bj{{\bf j}}
\def\bk{{\bf k}}

\def\bA{{\bf A}}
\def\bE{{\bf E}}

\def\bv{{\bf v}}
\def\bn{{\bf n}}

\def\bs{{\bf s}}
\def\nn{\nonumber}
\def\Acalv{{\boldsymbol {\mathcal A}}}
\def\Acal{{\cal A}}
\def\Bcalv{{\boldsymbol {\mathcal B}}}

\newcommand\commentout[1]{}

\def\be{\begin{equation}}
\def\ee{\end{equation}}
\def\ber{\begin{eqnarray}}
\def\eer{\end{eqnarray}}
\def\nn{\nonumber}
\def\bk{{\bf k}}

\newcommand{\ie}{{\it i.e.~}} 	
\newcommand{\eg}{{\it e.g.~}} 	

\definecolor{greenS}{rgb}{0.00, 0.6, 0.00}
\definecolor{orangeS}{rgb}{0.6, 0.1, 0.1}

\begin{document}
 \title{The spin Hall effect}
 \author{Cosimo Gorini}
\affiliation{SPEC, CEA, CNRS, Universit\'{e} Paris-Saclay, 91191 Gif-sur-Yvette, France}
 \date{\today}

\begin{abstract}
In metallic systems with spin-orbit coupling a longitudinal charge current may generate a transverse pure spin current; vice-versa an injected pure spin current may result in a transverse charge current.  Such direct and inverse spin Hall effects share the same microscopic origin: intrinsic band/device structure properties, external factors such as impurities, or a combination of both.  They allow all-electrical manipulation of the electronic spin degrees of freedom,\ie without magnetic elements, and their transverse nature makes them potentially dissipationless.  It is customary to talk of spin Hall effects in plural form, referring to a group of related phenomena typical of spin-orbit coupled systems of lowered symmetry.  
\end{abstract}


\maketitle


\begin{framed}
\subsection*{Key points/objectives}
\begin{itemize}
 \item History and definition of the effect(s)
 \item Phenomenology and concepts: spin-orbit coupling in solids, charge vs. spin currents, bulk vs. edge effects
 \item Experiments: from low- to room temperature, diversity and complexity of setups 
 \item Theory: the challenge of complexity, competition between different microscopic mechanisms, Onsager reciprocity 
 \item The broader context: generalisations of the spin Hall effect(s) and suggested further readings
\end{itemize}
\end{framed}

\section{Introduction}
\label{sec_intro}

\begin{figure}[h!]
 \includegraphics[width=.8\linewidth]{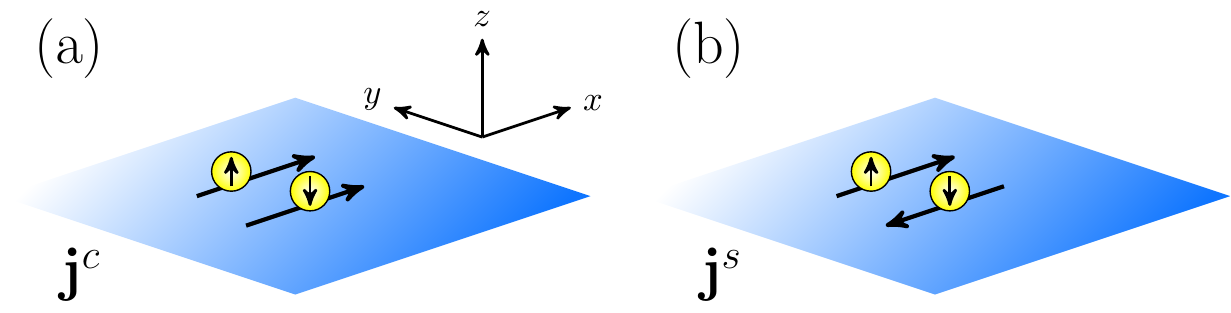}
 \caption{Left panel: A charge current ${\bf j}^c$ along $x$, \ie an overall spin-unpolarised ensemble of quasielectrons moving in the $y$ direction.  Right panel: A $z$-polarised pure spin current ${\bf j}^s$ flowing along $x$.}
 \label{fig_currents}
\end{figure}

A charge-carrying state in a metallic system is a (non-equilibrium) momentum-ordered state of an electronic ensemble: a collection of quasielectrons\footnote{I will restrict my discussion to systems where well-defined fermionic quasiparticles exist and make up a Fermi liquid.  Words such as ``quasielectron'' and ``electron'' or ``quasiparticle'' and ``particle'' will be used interchangeably.} moving preferentially in a given direction, see Fig.~\ref{fig_currents} (a).  However quasielectrons carry around their internal spin degrees of freedom, too.  For their ensemble to be spin carrying some form of spin order is needed.  There are two basic scenarios: 
\begin{itemize}
 \item Spin order is independent of momentum order.  This is the situation in a ferromagnet, where the spins of charge carriers align with the magnetisation via exchange coupling independently of their orbital motion.
 \item Spin and momentum orders are correlated.  This is possible in the presence of spin-orbit coupling, which quite generally creates correlations between orbital motion (momentum) and spin. 
\end{itemize}
Consider the second scenario, and to be definite the somewhat special situation sketched in Fig.~\ref{fig_currents} (b), where quasiparticles with opposite spin projection along $z$ move in opposite directions.  An ensemble of such particles does not carry any overall charge -- the associated charge current is zero, ${\bf j}^c=0$ -- but it does carry angular momentum: it is a pure spin-carrying state sustaining a ``pure spin current'' ${\bf j}^s\neq 0$ 
\ber
{\bf j}^c &\equiv& q\left({\bf j}^\uparrow + {\bf j}^\downarrow\right) = 0
\\ 
{\bf j}^s &\equiv& \frac{\hbar}{2}\left({\bf j}^\uparrow - {\bf j}^\downarrow\right) \neq 0, 
\eer
with $q$ the quasiparticle charge and $\hbar$ the reduced Planck constant.

\begin{figure}[h!]
 \includegraphics[width=0.8\linewidth]{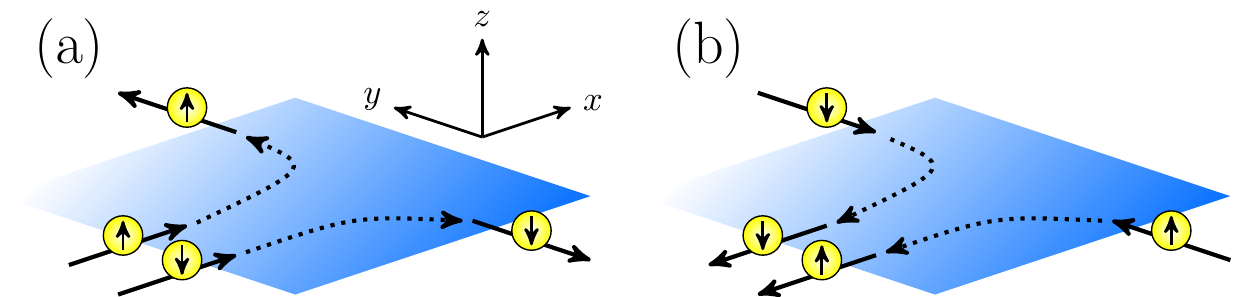}
 \caption{Left panel: spin Hall effect.  An $x$-flowing charge current is injected, and a $z$-polarised pure spin current along $y$ is generated via spin-orbit coupling.  Right panel: inverse spin Hall effect, where the role of the spin and charge currents are exchanged.  Note that in the sHe/isHe the charge current, spin current and the spin quantisation axis are all orthogonal to each other.}
 \label{fig_she_ishe}
\end{figure}

In the presence of spin-orbit coupling a transverse ${\bf j}^s$ can be generated from a longitudinal ${\bf j}^c$, or vice-versa, see Fig.~\ref{fig_she_ishe}.  These are respectively the spin Hall effect (sHe)\cite{dyakonov1971,hirsch1999} and its reciprocal version, the inverse spin Hall effect (isHe)\cite{dyakonovbook,engelbook}.  The sHe was first predicted by Dyakonov and Perel in 1971 \cite{dyakonov1971} but got his current name only in 1999, when Hirsch \cite{hirsch1999} sort of re-discovered it and its fortune really started.  The isHe followed a similar path.  
In the geometry from Fig.~\ref{fig_she_ishe} one has
\ber
\label{eq_SHE1}
j^s_y &=& (\hbar/q)\theta^{\rm sHe} j^c_x,
\\
\label{eq_ISHE1}
j^c_x &=& (q/\hbar)\theta^{\rm isHe} j^s_y
\eer
with $\theta^{\rm sHe}, \theta^{\rm isHe}$ the spin Hall and inverse spin Hall angles.  Note that from now on upper (lower) indices will denote charge/spin (real space) components. The spin Hall/inverse spin Hall angles are crucial quantities used as standard measure for the spin-charge conversion efficiency of a given setup \cite{valenzuela2006,hahn2013,obstbaum2014,sinova2015}.  In Secs.~\ref{sec_phenomenology} and \ref{sec_theory} we will show that their existence can be expected based on crude but general arguments, while their precise origin and values are extremely sensitive to the microscopic details of the system.  

The sHe and isHe exist also in quantised form.  Indeed, the ``quantum spin Hall insulator'' is the paradigmatic example of a time-reversal symmetric two-dimensional topologically insulating phase \cite{hasan2010}.  The latter is a phase of matter which is insulating in the bulk but hosts two perfectly conducting edge states which are time-reversed partners -- up electrons moving in one direction, down in the opposite.  In simple terms, a quantum spin Hall phase can be thought of as two time-reversed copies of a (time-reversal broken) quantum Hall state.  Its existence was confirmed experimentally in HgCdTe quantum wells in 2007 \cite{koenig2007}.  I will not discuss it further here, but refer the interested reader to the Encyclopedia Chapter ``Topological Effects''.   
 
The sHe and isHe are nowadays routinely employed in metallic systems to inject and/or read out spin signals via electrical means, and their technological relevance is rapidly on the rise since the early 2000s \cite{dyakonovbook,engelbook,sinova2015,fert2019}.  The rest of the Chapter will focus uniquely on them.  In particular I will not discuss the anomalous Hall effects, closely related phenomena which require however to break time-reversal symmetry \cite{dyakonovbook,nagaosa2010}.  The reader should indeed keep in mind that the spin Hall effect I will deal with in this Chapter actually belongs to a larger class of phenomena which may appear whenever quasiparticles with internal structure move in a non-trivial medium, \ie not just in the vacuum\footnote{In condensed matter one customarily talks of ``pseudospin'' internal degrees freedom, such as sublattice or valley pseudospin in graphene, which may give rise to different forms of ``pseudospin Hall effect''.}.
        
\section{Phenomenology and basic concepts}
\label{sec_phenomenology}

\subsection{The role of spin-orbit coupling}
Take an electron moving with velocity ${\bf v}$ in presence of an electric field ${\bf E}$.  In its reference frame the electron sees a velocity-dependent magnetic field
\be
{\bf B}'(\bv) = -\gamma\frac{{\bf v}}{c}\times{\bf E},\quad\gamma=\frac{1}{\sqrt{1-v^2/c^2}}
\ee
which couples to its spin $\bs=(\hbar/2)\bsigma$, with $\bsigma$ the vector of Pauli matrices, as usual
\be
\label{eq_so_0}
H_{\rm so} \sim \bs\cdot{\bf B}' \sim \bs\cdot\left(\bv\times\bE\right).
\ee
Here $c$ is the speed of light.  This is a primitive derivation of the spin-orbit interaction.  

\begin{figure}[h!]
 \includegraphics[width=0.8\linewidth]{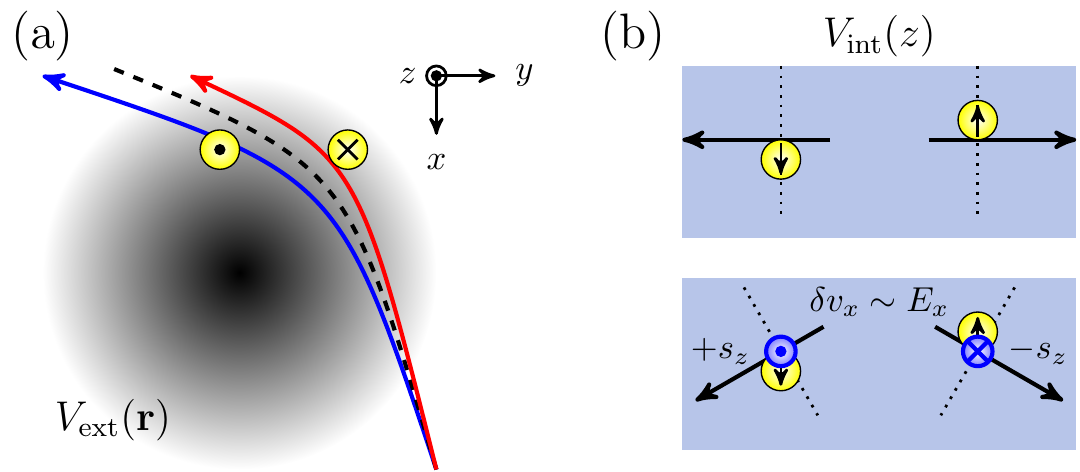}
 \caption{Left panel: Electrons with opposite spins are deflected differently by a scattering potential $V_{\rm ext}(\br)$. The dashed black line is the trajectory without spin-orbit corrections.  Right panel: Electrons confined to two dimensions by $V_{\rm int}(z)$.  The internal velocity-dependent field, see Eq.~\eqref{eq_Rashba_0}, defines the spin quantisation axis, shown with dotted lines.  The two electrons are time-reversed partners, and when driven by an electric field along $x$ their spins will precess around the new (tilted) quantisation axis, yielding out-of-plane components $\pm s_z$ shown in blue.  The result is a $z$-polarised spin current in the transverse $y$ direction.}
 \label{fig_scenarios}
\end{figure}

Let us assume that $\bE=-\nabla V_{\rm ext}$ comes from an external source, \eg the (random) electrostatic potential from an impurity, see Fig.~\ref{fig_scenarios} (a).  The quasiparticle spin thus couples to a non-homogeneous field $B'(\bv,\br)$, similarly to what happens in a Stern-Gerlach apparatus.  This has various consequences, \eg it causes spin-flip (Elliot-Yafet) relaxation, but in particular it results in a spin-dependent force ${\bf F} \sim - \nabla\left[\bs\cdot{\bf B}'\right]$ orthogonal to $\bv$   
\be
F_\perp^\uparrow = -F_\perp^\downarrow \sim -\nabla_\perp B'.
\ee   
This sideway separation of spin-up and spin-down quasiparticles is referred to as Mott skew scattering, and can split an incoming flux of unpolarised electrons into a transverse pure spin current, \ie it yields a finite $\theta^{\rm sHe}$.  It is an ``extrinsic'' mechanism requiring the presence of scattering centres \cite{dyakonovbook,hirsch1999,engelbook,culcer2010}, but a finite $\theta^{\rm sHe}$ may also have origins which are ``intrinsic'' to the device and/or band structure. 
Consider for example $\bE=-\nabla V_{\rm int}(z)$, with $V_{\rm int}(z)$ an internal potential which defines your structure, see Fig.~\ref{fig_scenarios} (b).  Such a potential breaks $z$ inversion symmetry and constrains electrons to move in the $x$-$y$ plane, as in semiconductor quantum wells.  One has
\be
\label{eq_Rashba_0}
H_{\rm so} \sim \bs \cdot \left( \bv \times e_z \right).
\ee  
Now $\bs$ couples to a homogeneous ${\bf B}'(\bv)$ which always points in plane, called a Rashba field \cite{bychkov1984,winklerbook}.  The eigenstates of the problem are spinors whose quantisation axis depends on their direction of motion.  As sketched in Fig.~\ref{fig_scenarios} (b), under an electrical bias one expects spin-momentum correlations to arise leading to a transverse, out-of-plane-polarised spin current -- that is, to a non-vanishing $\theta^{\rm sHe}$.  Recall that in this intrinsic example inversion symmetry was explicitly broken.  Some form of spatial symmetry-breaking is indeed a requirement of intrinsic scenarios beyond the present Rashba case \cite{engelbook,winklerbook}.  On the other hand time-reversal symmetry is preserved by spin-orbit interaction.  

Besides the canonical intrinsic and extrinsic mechanisms just described, other sources for the spin Hall effects include dynamical couplings with phonons \cite{gorini2015, karnad2018} or spin fluctuations \cite{okamoto2019}, the interplay of impurity and magnon scattering \cite{ohnuma2016}, or fluctuations of the Rashba field \cite{dugaev2012}.  Irrespectively of the origin, and just as a normal Hall current, the spin Hall current is transverse with respect to the drift velocity and thus potentially dissipationless.

\subsection{Size and form of (effective) spin-orbit coupling}
Since charge carriers in a metallic system move at the Fermi velocity $v_F\ll c$, should one expect $H_{\rm so}$ to be only a negligible relativistic correction?    
The answer is ``not necessarily''.  Electrons in solids do not move in the vacuum, but constantly get close to ionic cores from the lattice where unscreened very strong electric fields exist, compensating for their (relativistically) slow velocity.  One can indeed show that $s$-wave Bloch electrons close to the Fermi energy feel an effective spin-orbit interaction which reads
\be
\label{eq_effective_so1}
H_{\rm so} = \lambda\bsigma\times\nabla \delta V\cdot\bp,
\ee 
with $\delta V$ any potential other than that of the host lattice, \eg $\delta V=V_{\rm ext}$ or $\delta V=V_{\rm int}(z)$ from our previous examples.  While the form of $H_{\rm so}$ is the same as in the vacuum, the coupling constant $\lambda$ is an effective Compton wavelength strongly renormalised by the lattice: $\lambda/\lambda_0\sim 10^6$, with $\lambda_0$ the vacuum Compton wavelength.  The standard machinery behind such renormalization is $\bk\cdot\bp$ theory, which, together with some auxiliary techniques (L\"owding partitioning, Schrieffer-Wolff transformation,\dots), is just a systematic way of building low-energy models starting from the band structure (Bloch states) of a given lattice -- at the simplest level it leads \eg to the effective mass description of electrons in solids \cite{winklerbook,engelbook}.  Eq.~\eqref{eq_effective_so1} can actually be generalised to
\be
\label{eq_effective_so2}
H_{\rm so} = \bb(\bp)\cdot\bsigma,
\ee 
which looks like Zeeman coupling with a momentum-dependent internal field $\bb(\bp)$.  The latter defines a momentum space spin texture -- its vectorial image in reciprocal space -- of central importance well beyond spin Hall physics\footnote{In the presence of $\bb(\bp)$ the Hilbert space of the problem is usually equipped with a non-vanishing Berry curvature, with the potential for hosting non-trivial topological phases}.  This form of effective spin-orbit interaction is valid beyond the two example scenarios previously introduced, and is the starting point for most discussions of spin-orbit coupled transport phenomena \cite{engelbook,winklerbook}\footnote{There are situations in which an internal field $\bb(\bp)$ appears for microscopic reasons which have nothing to do with relativistic spin-orbit interaction.  The field thus couples to some internal pseudospin degree of freedom ${\boldsymbol \tau}$ of the low-energy quasiparticles, $\bb(\bp)\cdot{\boldsymbol\tau}$.  See comments in the closing of Sec.~\ref{sec_intro}.}.

\subsection{A closer look at spin currents}
The spin Hall/inverse spin Hall angles, see Eqs.~\eqref{eq_SHE1}, \eqref{eq_ISHE1}, are crucial quantities in spin Hall physics.  However to define them one needs to know exactly what a spin current is, which is not a completely trivial task.  Let us see why.
  
Charge is a conserved quantity respecting the continuity equation
\be
\partial_t n + \nabla\cdot\bj^c = 0,
\ee
with $n$ the charge (particle) density and $\bj^c = \bv n$ the current.  In formal terms, this is a consequence of gauge invariance: charge (particle number) is the conserved quantity associated with the $U(1)$ gauge symmetry of the system -- a case of Noether's theorem.  
     
A purely orbital Hamiltonian $H_0$ without spin-orbit interaction cannot affect the dynamics of the spin degrees of freedom: not only charge, but also spin is conserved for $H_0$.  Formally $H_0$ is spin-rotation invariant, which for spin 1/2 particles means that it has $SU(2)$ gauge symmetry.  Spin conservation yields the continuity equation  
\be
\label{eq_spin_continuity1}
\partial_t s^a + \nabla\cdot\bj^a = 0,\quad a = x,y,z,
\ee 
with $s^a$ the density of $a$-polarised electrons, and $\bj^a = \bv s^a$ the corresponding $a$-spin current density.   

In the presence of spin-orbit coupling, $H_0 \to H = H_0 + H_{\rm so}$, spin rotation symmetry is broken and spin is not conserved anymore.  Consider first extrinsic spin-orbit due to diluted impurities.  In this case spin is not conserved during scattering events, but remains so between them.  The definition $\bj^a = \bv s^a$ is thus good during flight, but at each scattering the $a$-polarised flux may rotate, partially split and lose weight due to spin-flip scattering.  The result is a spin relaxation rate $1/\tau^a_s$, with $\tau^a_s$ the $a$-spin lifetime, 
and some further extrinsic spin torque $\Gamma^a_{\rm ext}$
\be
\label{eq_spin_continuity2}
\partial_t s^a + \nabla\cdot\bj^a = -\frac{s^a}{\tau^a_s} + \Gamma^a_{\rm ext}.
\ee
The extra torque describes spin non-conservation effects beyond simple relaxation, \eg skew-scattering physics, which may directly couple spin and charge degrees of freedom.

If intrinsic spin-orbit coupling is present things are more complicated.  First, in this case the velocity $\bv = \partial_\bp H$ does not in general commute with the spin operator, which is obvious looking at Eq.~\eqref{eq_effective_so2}.  One can still generalise the spin current definition by symmetrization
\be
\label{eq_spin_current1}
\bj^a = \frac{1}{2}\left\{\bv,s^a\right\},
\ee
with $\left\{A,B\right\} = AB + BA$ the anticommutator.  However spin is now intrinsically -- \ie everywhere and always -- not conserved, {\it ergo} the continuity equation obeyed by such a current must be modified by an intrinsic spin torque $\Gamma^a_{\rm int}$
\be
\label{eq_spin_continuity3}
\partial_t s^a + \nabla\cdot\bj^a = \Gamma^a_{\rm int}.
\ee
The precise form of the torque is fixed by the internal spin-orbit field \eqref{eq_effective_so2}.  The intrinsic lack of a conservation law means that $\bj^a$, and thus $\Gamma^a$, are not uniquely defined now.  This is no fundamental problem, in the sense that not all physically meaningful currents need be conserved.  It can however be a delicate operative problem, as changing the definition of $\bj^a$ will change the definitions of the spin Hall/inverse spin Hall angles, which is what experiments typically aim at measuring.  Similarly, it will modify the estimates for the spin accumulations generated by the current, the accumulation being another popular observable.  Indeed, it is not alway obvious how (if) local spin currents somewhere, say in the bulk, are connected with local spin accumulations elsewhere, \eg at the sample's edges \cite{dyakonovbook,sonin2007,adagideli2007,sonin2010,gorini2012}.  These issues are recurringly discussed in the specialised literature \cite{rashba2003,nikolic2006,shi2006,sugimoto2006,sonin2007,adagideli2007,tokatly2008, sonin2010,sonin2010b,gorini2012,khaetskii2014} and can be dealt with in different ways, for example:
\begin{itemize}
 \item One can avoid referring to spin currents within the spin-orbit coupled region, and define them only in the metallic electrodes attached to sample, where $H_{\rm so}$ is negligible.  This is the picture naturally arising in the Landauer-B\"uttiker approach to transport \cite{jacquod2012}.  It is often more or less implicitly assumed in phenomenological discussions of experiments.
 \item One can focus on spin accumulations, \ie consider the equation of motion for the spin density without assuming a given form for the spin currents.  The appropriate form of the currents may be derived from the density equations, notably in the diffusive regime \cite{burkov2004,malshukov2005,raimondi2006,adagideli2007}.
 \item One can use the non-Abelian gauge properties of the Hamiltonian $H$, \ie its spin rotation $[SU(2)]$ properties, to define spin currents much as colour currents are defined in high-energy physics \cite{tokatly2008,gorini2010}.  This approach removes any ambiguity from the definition of $\bj^a, \Gamma^a$, but cannot directly be exploited for any form of ${\bb}(\bp)$.  I will comment further on it in Sec.~\ref{sec_theory}.
 \item One can try to define a conserved spin current by combining $\bj^a$ and $\Gamma^a$ \cite{shi2006,sugimoto2006}. 
\end{itemize}
There is arguably no single ``best'' approach.  One should decide which way to go based on the specific physical situation, and -- if the physics allow -- on personal tastes.  It should also be kept in mind that both extrinsic and intrinsic processes are present in typical setups, and that they may yield further intrinsic-extrinsic crossed processes, \ie the corresponding torques are not simply additive \cite{raimondi2009,raimondi2012}.  Moreover -- and independently of the spin current definition -- continuity equations like \eqref{eq_spin_continuity1}, \eqref{eq_spin_continuity2} or \eqref{eq_spin_continuity3} must be supplemented with appropriate boundary conditions \eg at interfaces between different materials or at the egdes of the system, which may yield additional (local) torques \cite{adagideli2005,raimondi2006,adagideli2007,tserkovnyak2007,malshukov2007, khaetskii2014,amin2016,toelle2018}.  

\section{Experiments}
\label{sec_experiments}

The first spin Hall experiments were performed in the 1970s and 1980s in semiconductors \cite{dyakonovbook}, but relatively few got interested at the time.  The business became fashionable in the early 2000s, and the spin Hall effects are nowadays not only the object of fundamental research \cite{zhu2019,nakagawara2019,li2021,yanez2021,boventer2021}, but also established tools in more application-oriented settings \cite{liu2012,fert2019,safeer2020,karube2020}.    In fact, while the first experiments were performed at fairly low temperatures (a few to a few tens of Kelvins) room temperature measurements are routine today, and large spin Hall angles have been reported in different materials.  To give a rough idea of the progression, the spin Hall angle reported in a pioneering experiment by Valenzuela and Tinkham in 2006 was $\theta^{\rm sHe} \sim 10^{-4}$ in Al at T=4.2K, while less than 10 years later room temperature measurements in Pt, Ta or W reached $\theta^{\rm sHe} \sim 10^{-1}$ \cite{obstbaum2014,rojassanchez2014,liu2012,hahn2013,pai2012}.

\begin{figure}[h!]
 \includegraphics[width=0.8\linewidth]{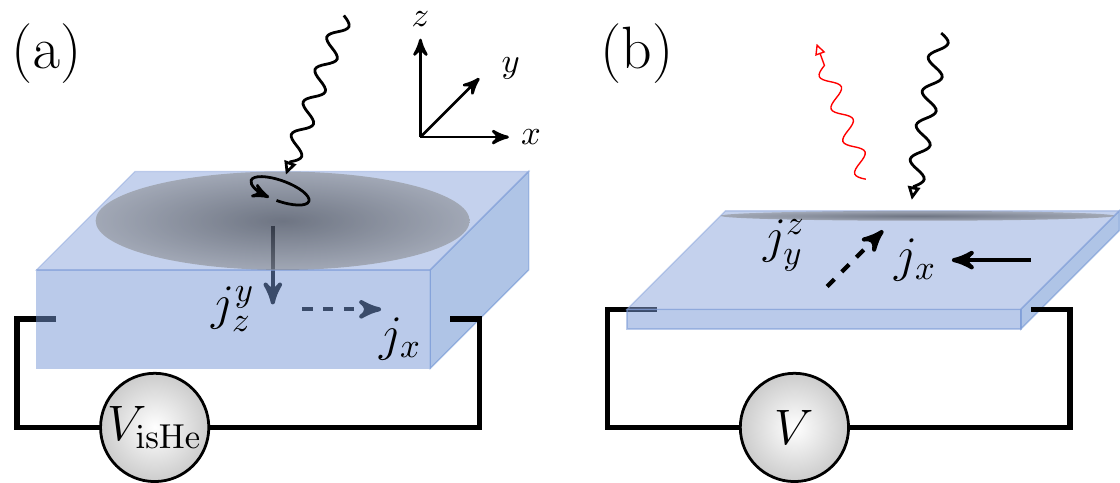}
 \caption{Left panel: A broad spot of circularly polarised light creates a non-equilibrium spin polarisation $\delta s^y$ at a sample's surface (shaded region), and a a diffusion spin current $j^y_z \sim \partial_z \delta s^y$ flows into the 3D bulk.  Here the isHe generates a transverse charge current $j_x$, yielding a finite output voltage $V_{\rm isHe}$.  Setups of this kind were proposed already in the early days of spin Hall physics \cite{dyakonovbook}.  Right panel: In a 2D system the charge current $j_x$ from an applied bias $V$ is converted into a spin Hall current $j^z_y$.  The resulting spin accumulation at the edges (shaded region) is measured by Kerr rotation of scattered light (red).  The technique was employed in the first experimental observation of the sHe in a 2D electron gas \cite{kato2004}, and a similar one in a 2D hole gas \cite{wunderlich2005}.  The connection between bulk spin currents and edge spin accumulations can however be less direct than this cartoon suggests \cite{dyakonovbook,nikolic2006,adagideli2007,sonin2010,gorini2012}.}
 \label{fig_optical_setups}
\end{figure}

\begin{figure}[h!]
 \includegraphics[width=0.8\linewidth]{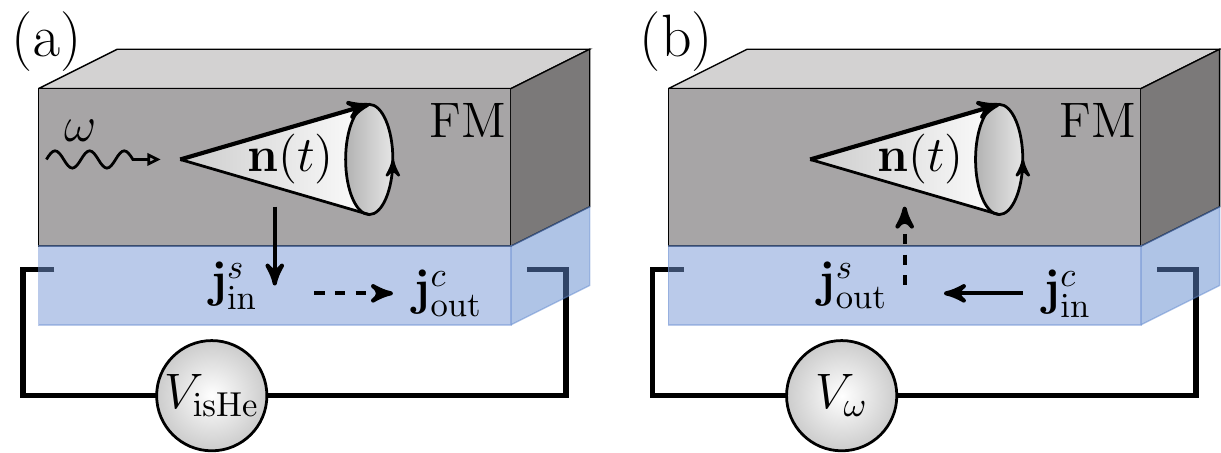}
 \caption{Left panel: Sketch of a spin pumping setup.  Microwaves in a ferro-/ferrimagnet FM at frequency $\omega$ drive the magnetisation, ${\bf M}=M\bn\to{{\bf M}(t)=M\bn(t)}$.  Its precession injects angular momentum into the underlying spin-orbit coupled normal metal, generating a spin current $\bj^s$.  The latter is converted by the isHe into a charge current $\bj^c$ and {\it ergo} a measurable voltage $V_{\rm isHe}$, in general with both DC and AC components \cite{jiao2013,wei2014,weiler2014}.  Right panel: A reverse-bias scenario, in which a charge current at frequency $\omega$ is injected and converted by the sHe into an AC spin current.  The latter exerts a torque on ${\bf M}$ and drives its precession.  There are numerous DC and AC variations to these schemes, involving many different magneto-resistive effects modulated by spin-orbit interaction, see Refs.~[\onlinecite{sinova2015,fert2019}] for an overview.}
 \label{fig_magneto-electric_setups}
\end{figure}

\begin{figure}[h!]
 \includegraphics[width=0.8\linewidth]{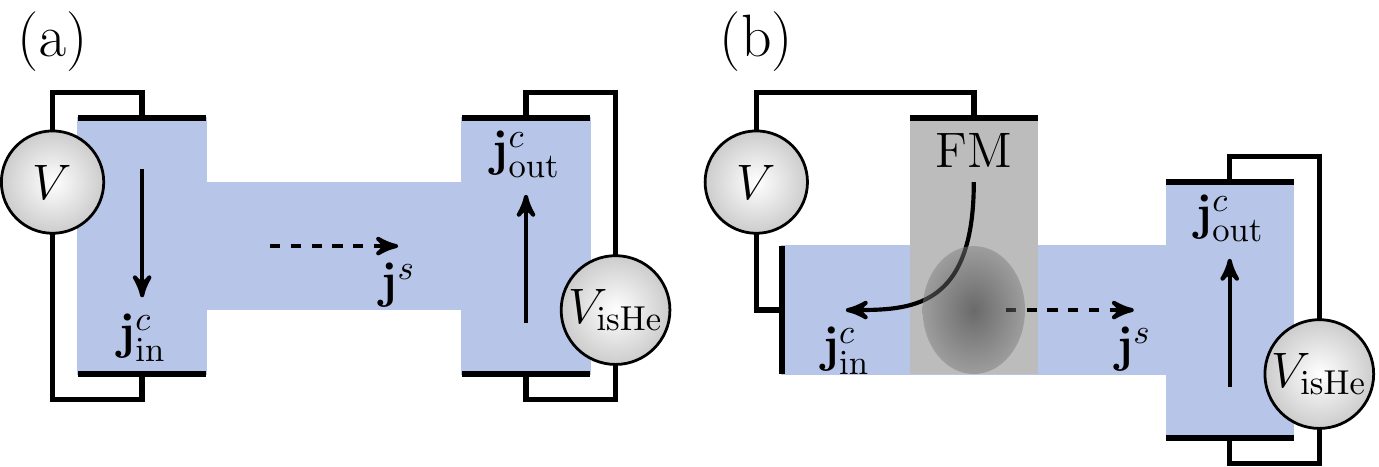}
 \caption{Left panel: Hall bar geometry without magnetic elements.  The injected charge current $\bj^c_{\rm in}$ is converted into a transverse spin current $\bj_s$ by the sHe.  The latter is converted back into a charge current $\bj^c_{\rm out}$ by the isHe in the right arm of the setup, yielding a finite $V_{\rm isHe}$.  One of the earliest implementations of this setup was used to measure the ballistic sHe in a HgTe quantum well \cite{bruene2010}.  Right panel: non-local setup with a ferromagnetic (FM) electrode, shown in dark grey, deposited on top of a T-shaped metallic film.  The injected current $\bj^c_{\rm in}$ is drained to the left contact, since the right metallic arm is at the same electrochemical potential as the FM electrode.  The current $\bj^c_{\rm in}$ is spin polarised, therefore it creates a non-equilibrium spin accumulation underneath the FM contact, shown by the shaded area.  Part of it diffuses towards the right, yielding a pure spin current $\bj^s$ which is then converted by the isHe into a measurable transverse voltage $V_{\rm isHe}$.  The scheme was first employed by Valenzuela and Tinkham \cite{valenzuela2006}.}
 \label{fig_all-electrical_setups}
\end{figure}

The numerous experimental schemes available can roughly be divided into three classes.
\begin{itemize}
\item Optical setups, historically the first to be used to detect the sHe/isHe, exploit the interaction between polarised light and the spin of charge carriers.  For example, circularly polarised light can be used to generate local non-equilibrium spin accumulations which later diffuse through the system.  The diffusion spin current can then be converted by the isHe into charge signals measured with standard electrodes, see Fig.~\ref{fig_optical_setups} (a).  On the other hand, spin accumulations can be measured by circularly polarised electroluminescence or magneto-optical Kerr and Faraday effects.  An example is shown in Fig.~\ref{fig_optical_setups} (b), where the spin accumulation at the edge of the system generated by the sHe is measured by the degree of (Kerr) rotation of light scattered off the sample.  All-optical schemes are also employed \cite{werake2011,seifert2016}, allowing in particular time-resolved experiments on ultra-short (THz) timescales \cite{werake2011} -- which are not accessible to electronic systems.  
\item Magneto-electric (spin pumping, spin torque) setups, relying on the interplay between magnetisation and spin dynamics \cite{tserkovnyak2005}.  Fig.~\ref{fig_magneto-electric_setups} shows a paradigmatic example: a spin current $\bj^s_{\rm in}$ is injected from an out-of-equilibrium magnetic element, \eg a (conducting or insulating) magnet driven by microwaves, and converted into a charge current $\bj^c_{\rm out}$ collected at a normal metallic electrode.  The latter can actually be used to run the experiment under ``reverse bias'': $\bj^c_{\rm in}$ is injected, and one measures the torque that the resulting $\bj^s_{\rm out}$ exerts on the adjacent magnet.  The presence of magnetic elements considerably increases the degree of complexity of the overall system, and may lead to additional effects which are however beyond the scope of this short overview \cite{tserkovnyak2005,manchon2019}.  Time-dependent experiments are also performed, \eg to measure AC spin Hall effects \cite{hahn2013b,wei2014,weiler2014}.
\item All-electrical setups, conceptually probably the simplest.  Fig.~\ref{fig_all-electrical_setups} (a) shows a most basic one, without any magnetic element: A charge current $\bj^c_{\rm in}$ is injected by a metallic electrode, is converted into a spin signal by the sHe, and finally re-converted by the isHe into an outgoing $\bj^c_{\rm out}$ collected at some other electrode.  A very popular scheme requiring a magnetic electrode is instead sketched in Fig.~\ref{fig_all-electrical_setups} (b).  Both are non-local -- input electrodes are somewhere, output electrodes elsewhere -- a common feature in spin Hall setups \cite{valenzuela2006,takahashi2007}. 
\end{itemize}  

The boundary between classes is clearly blurred, and mixed techniques are often employed.  An important general observation is that experimental setups -- apart perhaps from the simplest all-electrical ones -- are fairly complex, consisting of multiple elements of different nature subject to various kinds of drivings.  Paired with the vast number of spin-charge (or charge-spin) conversion channels available in any given system, this makes for interesting debates concerning possible microscopic interpretations of experiments\footnote{The spin galvanic and inverse spin galvanic effects are very often crucial ``partners'' of the spin Hall effects \cite{ganichev2002,ganichev2016,shen2014,gorini2017}.  They are another common channel of spin-charge/charge-spin conversion, discussed in detail in a dedicated Encyclopedia Chapter}.

\section{Theory}
\label{sec_theory}

\subsection{An instructive example and the general framework}
The spin Hall effects are non-equilibrium phenomena handled with the usual arsenal of transport theory techniques: Keldysh formalism, density matrix and semiclassical kinetics, Kubo formula, Landauer-B\"uttiker formalism \dots.  Irrespective of the techniques employed, a source of substantial theoretical challenges is complexity.  In crude terms, the Hamiltonian of a spin Hall system requires numerous ingredients, recall the discussion from Sec.~\ref{sec_experiments}.  The resulting quasiparticle dynamics are in general quite sensitive to the presence/absence of, and competition between, each.

To see this in a concrete way it is instructive to start from the barebone model of an ideal Rashba system
\be
\label{eq_Rashba_1}
H_{\rm R} = \frac{p^2}{2m} + \frac{\alpha}{\hbar}\left[p_y\sigma^x - p_x\sigma^y\right],
\ee
where $m$ is the effective electron mass and $\alpha \sim \nabla V_{\rm int}(z)$ the Rashba coupling constant, proportional to the (effective) electric field confining the electrons to the $x$-$y$ plane -- see the heuristic discussion in Sec.~\ref{sec_phenomenology}, Eq.~\eqref{eq_Rashba_0}.

Given $H_R$, the goal is to compute the frequency-dependent spin Hall conductivity $\sigma^{\rm sHe}(\omega)$, defined as
\be
j^z_y(\omega) = \sigma^{\rm sHe}(\omega) E_x(\omega),
\ee
with $j^z_y$ the $z$-polarised spin current in the $y$ direction.  The standard choice is to take the symmetrised spin current definition \eqref{eq_spin_current1} -- other choices are possible, recall the discussion from Sec.~\ref{sec_phenomenology}, and the consequences will be addressed below.  The spin Hall conductivity can be written in terms of the spin current-charge current Kubo response function $\langle\langle j^z_y ; j_x \rangle\rangle_\omega \equiv -\frac{i}{\hbar}\int_0^t \left[j^z_y(t),j_x(0)\right]e^{i\omega t}{\rm d}t$, with $[A,B]= AB-BA$ the commutator \cite{landaubook}.  Since the charge current couples to the vector potential as $\bj\cdot\bA$, and $E_x(\omega)=-i\omega A_x(\omega)$, one has
\be
\sigma^{\rm sHe}(\omega) = \frac{\langle\langle j^z_y ; j_x \rangle\rangle_\omega}{i\omega}.
\ee 
Once the DC $\sigma^{\rm sHe}\equiv\lim_{\omega\to 0}\sigma^{\rm sHe}(\omega)$ is known the spin Hall angle follows
\be
\theta^{\rm sHe} = \frac{q}{\hbar} \frac{\sigma^{\rm sHe}}{\sigma^c_x},
\ee
with $\sigma^c_x$ the longitudinal DC charge conductivity.  An explicit computation yields the ``universal'' DC result
\be
\label{SHE_clean}
\sigma^{\rm sHe}_{\rm clean} = \frac{e}{8\pi\hbar},
\ee
with the electron charge $q = -e < 0$.
The subscript ``clean'' highlights that the system is without any defects.  Such a beautiful result, due to the intrinsic Berry phase of electrons on the Rashba Fermi surface, is unfortunately very fragile.  If one adds dirt to the model, \ie a random impurity potential, $H_{\rm R}\to H_{\rm R} + V_{\rm imp}(\br)$, the spin Hall conductivity exactly vanishes
\be
\label{SHE_dirty}
\sigma^{\rm sHe}_{\rm dirty} = 0.
\ee
The vanishing is diagrammatically subtle: since it comes from vertex corrections, it cannot be guessed by simply introducing a disorder broadening of the momentum eigenstates in the Kubo response kernel \cite{schwab2002,inoue2004,raimondi2005}.  Indeed, it was overlooked at first in the scientific literature \cite{sinova2004}.  On the other hand, it is easily understood with kinetic arguments \cite{dimitrova2005,chalaev2005}, since the homogeneous continuity equation for the $y$-spin component reads
\be
\partial_t s^y = \underbrace{-\frac{2m\alpha}{\hbar^2} j^z_y}_{\Gamma^y_{\rm int}}.
\ee
At steady state the spin current $j^z_y=0$.

Eq.~\eqref{SHE_dirty} is as fragile as its clean counterpart \eqref{SHE_clean}.  If one further adds extrinsic spin-orbit interaction, that is spin-orbit interaction with the impurity potential, $H_{\rm R}\to H_{\rm R} + V_{\rm imp}(\br) + \lambda \bsigma\times\nabla V_{\rm imp}(\br)\cdot\bp$, the spin Hall conductivity is non-zero
\be
\sigma^{\rm sHe}_{\rm int + ext} \neq 0,
\ee  
and furthermore depends non-trivially on different system parameters.  In particular \cite{raimondi2009,raimondi2012}
\be
\sigma^{\rm sHe}_{\rm int + ext} \neq \sigma^{\rm sHe}_{\rm int} + \sigma^{\rm sHe}_{\rm ext}.
\ee
Equivalent results would have been reached starting from the (linear) Dresselhaus model
\be
H_D = \frac{p^2}{2m} + \frac{\beta}{\hbar}\left[p_x\sigma^x - p_y\sigma^y\right],
\ee
where the coupling constant $\beta$ is now due to bulk inversion asymmetry, \ie the lack of inversion symmetry of the underlying crystal, as in zincblend compounds \cite{winklerbook}.  

The lesson to be learned from this example is not that low-energy effective models of Rashba or Dresselhaus type are unreliable -- quite the contrary, they are pillars of spintronics, even if more complex models are often needed \eg for precise quantitative comparisons with experiments.  It is rather that any effect crucially depending on the coupling between orbital motion and internal (spin) dynamics is subtler than standard charge-only transport phenomena, even when their description is based on the simplest models.  As a corollary, attacking the problem from different angles -- Kubo {\it vs.} kinetics in this case -- can be a good idea.   

The Rashba and Dresselhaus scenarios just considered are examples of a standard approach to transport widely employed throughout condensed matter.  The latter starts from some low-energy ($\bk\cdot\bp$) effective model whose parameters can be computed with {\it ab-initio} methods, or left as symmetry-allowed parameters to be estimated by comparison with experiments.  In our case the minimal Hamiltonian for spin 1/2 quasiparticles reads
\be
\label{eq_effective_H3}
H = H_0 + \bb(\bp)\cdot\bsigma + \delta H
\ee  
where $H_0$ describes band-bottom (top) free electrons (holes), and $\bb(\bp)$ is the effective intrinsic spin-orbit field, see Eq.~\eqref{eq_effective_so2}.  Higher-dimensional models (4 x 4, 6 x 6 \dots) are employed whenever more than a single $s$-band lie close to the Fermi energy, which is the case \eg for graphene \cite{castroneto2009,kochan2017,dyrdal2012,milletari2017}, Pt \cite{guo2008} or typically for holes \cite{winklerbook,engelbook}.  On the other hand there are situations in which the term $\bb(\bp)$ is negligible, \eg in bulk Al or Cu.  The last term $\delta H$ contains all extra ingredients needed in the specific situation, \eg exchange coupling with a magnetic texture, extrinsic spin-orbit coupling, disorder, phonons and so on.

The effective Hamiltonian \eqref{eq_effective_H3} is used to study non-equilibrium dynamics with whatever analytical and/or numerical techniques one prefers.  The approach is thus very general and flexible.  Alternatively, it is also possible to stick to {\it ab-initio} methods and use an atomistic Hamiltonian throughout.  In this case one typically relies on Kubo linear response formalism to compute the relevant transport coefficients, \eg $\sigma^{\rm sHe}$ \cite{lowitzer2011,koedderitzsch2015}.  

\subsection{Onsager reciprocity and a non-Abelian gauge field point of view}    
The sHe and isHe connect spin currents, even under time-reversal, with charge currents, odd under time-reversal.  From the general properties of Kubo response functions \cite{landaubook} there follows 
\be
\langle\langle j^z_y; j_x \rangle\rangle_\omega = - \langle\langle j_x; j^z_y \rangle\rangle_\omega,
\ee  
which implies
\be
\sigma^{\rm sHe} = -\sigma^{\rm isHe}.
\ee
This is the (linear response) Onsager relation between the sHe and isHe.  It is evident that changing the definition of the spin current changes the value of $\sigma^{\rm sHe}$ and $\sigma^{\rm isHe}$.  This is critical if a direct comparison with experiments is seeked: what spin current is being excited in the experimental setup?  What spin Hall angle are we talking about?  As discussed in Sec.~\eqref{sec_phenomenology} there are numerous ways to remove any ambiguity from such a comparison.  In particular, if one is not interested in local quantities such as conductivities, the problem can be bypassed by considering conductances between metallic leads without spin-orbit interaction \cite{jacquod2012}.  On the other hand a change of spin current definition does not break Onsager reciprocity if done consistently, \ie if the same definition is used to describe both the direct, say sHe, and reverse bias, say isHe, scenario.   

Another source of concern in the early 2000s was that the standard definition of spin currents, Eq.~\eqref{eq_spin_current1}, may yield non-vanishing equilibrium (circulating) currents \cite{rashba2003}.  Different authors highlighted however that there is nothing intrinsically unphysical or surprising in this \cite{sonin2007b,tokatly2008}: spin currents are even under time-reversal, so can exist in equilibrium, and physical systems hosting different kinds of equilibrium currents anyway exist \cite{heurich2003,sonin2007b,tokatly2008,sonin2010b}.  Indeed, adopting a non-Abelian gauge field point of view, equilibrium spin currents can be identified with the non-Abelian analogous of dissipationless Landau paramagnetic currents in solids \cite{tokatly2008}.

The non-Abelian gauge field approach requires to rewrite the spin-orbit interaction in terms of a non-Abelian vector potential $\Acalv$, \ie a tensor $\Acal^a_i$ with both spin $(a)$ and real space $(i)$ indices.  To be definite, for the Rashba Hamiltonian \eqref{eq_Rashba_1} one has
\be
\label{eq_rewriting}
H_R = \frac{\left(\bp+\Acalv\right)^2}{2m} + {\rm const.} 
\ee
with $\Acal^y_x = -\Acal^x_y = 2m\alpha/\hbar^2$, while $\Acal^a_j = 0$ for all other components.  The spin current immediately follows from $j^a_i = \partial H_R/\partial \Acal^a_i$.  It coincides with the standard definition \eqref{eq_spin_current1} and generally consists of both transport contributions and a non-dissipative equilibrium part.  Pursuing this route \eg in a diffusive sample, one obtains in particular a clear parallel between the standard Hall current $\bj_{\rm Hall}$ in presence of a magnetic field ${\bf B}$ and the $a$-polarised spin Hall current $\bj^a_{\rm sHe}$ in presence of a non-Abelian pseudomagnetic field $\Bcalv$ generated by $\Acalv$ \cite{gorini2010}
\be
\bj_{\rm Hall} = \frac{q\tau}{m} \bj \times {\bf B}
\; \to \;
\bj^a_{\rm sHe} = \frac{q\tau}{4m} \bj \times \Bcalv^a.
\ee       

The non-Abelian gauge field approach is based on relatively old ideas \cite{mathur1992,froehlich1993,brouwer2002}, but was recently revived to describe spin-charge coupled transport in different settings \cite{gorini2010,raimondi2012,adagideli2012,bergeret2013,tokatly2016,toelle2017, bobkova2017,jacobsen2017,malshukov2017,aikebaier2019,koenig2021}.  While its merits are evident, one should realise that a rewriting like Eq.~\eqref{eq_rewriting} is not always possible.  I refer to the relevant literature for details.

\section{Conclusions}
\label{sec_conclusions}

The spin Hall effects are a family of transverse transport phenomena appearing in (pseudo)spin-orbit coupled systems.  A good chunk of the theory and experimental background was established in the 1970s-1980s, but the effects became widely known in condensed matter only starting from the early 2000s, and are nowadays cornerstones of both fundamental and applied spintronic research.  Such a late blooming is probably due in good part to two roughly contemporary events.  First, the widespread realisation of the importance of geometry/topology-related concepts for Bloch electrons.  Since the latter usually require quasiparticles to have an internal structure, this strongly increased interest for (pseudo)spin-orbit coupled dynamics.  Second, technological advances which notably allowed the fabrication of high-quality semiconductor heterostructures, where spin manipulation became possible with a high level of precision, soon after followed by the discovery and functionalisation of graphene and other materials with strong (pseudo)spin-orbit interaction.

In this short overview I focused on the core, standard forms of the spin Hall effects, as they exist in normal Fermi liquids.  In this context they are active in a wide range of parameters, from large samples at room temperature -- important for potential applications -- down to mesoscopic samples at low temperatures.  However they may also appear in \eg strongly disordered systems \cite{smirnov2017}, superconductors \cite{takahashi2011}, metallic antiferromagnets \cite{zhang2014,gulbrandsen2020}, as ``valley Hall effects'' in different materials \cite{gorbachev2014,mak2014,lensky2015} or in the propagation of magnons \cite{onose2010,mook2014} and light \cite{hosten2008}.  They may also contribute to other transport effects, such as the spin Hall magnetoresistance \cite{chen2016,oyanagi2021,chen2013}.  In short, they are potentially present in any scenario where transport is due to quasiparticles with some internal structure which couples to a non-trivial background.     
 
\subsection{Notes on further readings}
The literature on the spin Hall effects is vast and ramifies quickly to neighbouring subfields.  The bibliography given here is meant to provide barebone directions to the newcomer, but is by no means exhaustive.  Numerous review articles, each with its own qualities and shortcomings, are available to the interested reader.  Refs.~[\onlinecite{dyakonovbook}] and [\onlinecite{engelbook}] are both must-read works.  Ref.~[\onlinecite{dyakonovbook}] provides in particular a thorough historical overview and excellent phenomenological discussions, while Ref.~[\onlinecite{engelbook}] offers a high-quality and very compact introduction to the technical background, introducing also modern topological concepts.  I also suggest Ref.~[\onlinecite{fert2019}] for a recent, short and more application-oriented discussion, and Ref.~[\onlinecite{vignale2010}] for its theory part.  Finally, Ref.~[\onlinecite{sinova2015}] provides an excellent experimental overview.

\section{Acknowledgements}
I am indebted to Lin Chen for a careful reading of the manuscript, and to Leonid E. Golub for useful comments.  I also thank all STherQO members for discussions, in particular Guillaume Weick and R\'{e}my Dubertrand.

\bibliography{SGE_biblio}

\end{document}